\documentclass[aps,twocolumn,floats,prl]{revtex4}
\usepackage{graphics,graphicx,epsfig}
\usepackage{amssymb,color}
\usepackage{epsf,epstopdf,wrapfig}
\usepackage[normalem]{ulem}

\newcommand{\beq}{\begin{equation}}
\newcommand{\eeq}{\end{equation}}
\newcommand{\beqn}{\begin{eqnarray}}
\newcommand{\eeqn}{\end{eqnarray}}
\newcommand {\e}[1]{\mathrm{~#1}}    
\newcommand {\E}[1]{\cdot 10^{#1}}		
\begin{document}

\title{Retinal metric: a stimulus distance measure derived from population neural responses}

\author{Ga\v{s}per Tka\v{c}ik\footnote{gtkacik@ist.ac.at}$^{a}$, Einat Granot-Atedgi$^{b}$, Ronen Segev$^{c}$
  and Elad Schneidman\footnote{elad.schneidman@weizmann.ac.il}$^b$}

\affiliation{$^a$Institute of Science and Technology Austria, Am Campus 1, A-3400 Klosterneuburg, Austria\\
$^b$Department of Neurobiology, Weizmann Institute of Science, 76100 Rehovot, Israel\\
$^c$Faculty of Natural Sciences, Department of Life Sciences and Zlotowski Center for Neuroscience, Ben Gurion University of the Negev, 84105 Be'er Sheva, Israel}

\date{\today}

\begin{abstract}
The ability of the organism to distinguish between various stimuli is limited by the structure and noise in the population code of its sensory neurons. 
Here we infer a distance measure on the stimulus space directly from the recorded activity of 100 neurons in the salamander retina. In contrast to previously used measures of stimulus similarity,  this ``neural metric'' tells us how distinguishable a pair of stimulus clips is to the retina, given the noise in the neural population response. We show that the retinal distance strongly deviates  from Euclidean, or any static metric, yet has a simple structure: we identify the stimulus features that the neural population is jointly sensitive to, and show the SVM-like kernel function relating the stimulus and neural response spaces. We show that the non-Euclidean nature of the retinal distance has important consequences for neural decoding.

\end{abstract}

\pacs{PACS numbers: 87.18.Sn, 87.19.Dd, 89.70.+c}

\maketitle


Neural populations convey information about their stimuli by their joint spiking patterns \cite{spikes}. 
At the level of single cells, the mapping from stimuli to spikes has often been captured by linear-nonlinear (LN) models \cite{sta,schwartz06}. 
Geometrically, a single LN neuron can  be viewed as a {\em perceptron} \cite{rosenblatt58}, partitioning the stimulus space into two domains---one of stimuli that  evoke spikes and one of stimuli  that do not---by a decision boundary, or a hyperplane, determined by the linear feature of the LN model \cite{rdr0,baa1,rdr}. The brain, 
listening for spikes coming from such a neuron, will thus interpret stimuli as similar insofar as they evoke similar spiking patterns.
But how does an interacting population, as a whole,  partition the stimulus space? Conversely, which stimuli are interpreted as different, or similar, by an interacting population?

Answering these  questions is fundamental to our understanding of the neural code and  depends critically on  finding the correct ``metric'' for sensory stimuli in terms of the information that neural populations carry. 
Since neurons are noisy, repeated presentations of the same stimulus can result in different neural responses,  so the stimulus/response mapping of the population needs to be described by the probability distribution, $P(\sigma|s)$ \cite{rdr97}. Two  stimuli $s_1$ and $s_2$ may be far apart as measured by a chosen  distance function, e.g. Euclidean norm $D_2(s_1,s_2)=||s_1-s_2||$, yet they could evoke responses drawn from almost overlapping distributions $P(\sigma|s_1)$ and $P(\sigma|s_2)$, making it nearly impossible for the brain, listening  to the spikes $\sigma$ arriving from the sensory system, to tell those stimuli apart. Conversely, the sensory circuit could  be  sensitive to specific  stimulus changes that have a small Euclidean norm, emphasizing those particular differences as an important feature and encoding it in the neural response. We therefore suggest that the biologically relevant  distance between pairs of stimuli  should be derived from the distance between the response distributions evoked by these stimuli \cite{shepard57,yy92}.

\begin{figure*} 
\centering
\includegraphics[width=6.5in]{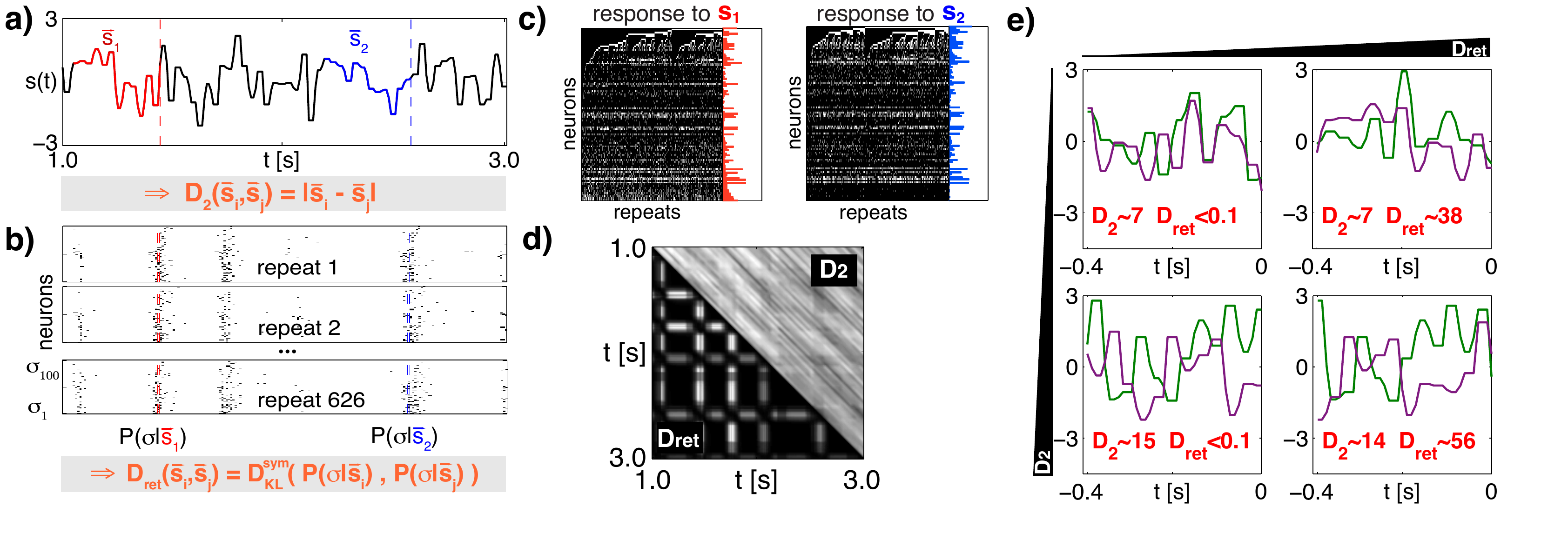} 
\caption{
{\bf a)} Stimulus segment with two 400 ms stimulus clips $s_1$ (red), $s_2$ (blue). 
 {\bf b)} Rasters for 100 RGCs shown for 3 example repeats; response vectors to $s_1$, $s_2$ shown between dashed  lines. 
 {\bf c)} Measured response rasters to two highlighted stimulus clips ($s_1$, left; $s_2$, right; spikes = white dots; colored bars = neuron firing rates). 
{\bf d)} For every pair of timebins in the experiment, the Euclidean distance $D_2$  between the corresponding stimulus clips is shown in the upper diagonal part of the matrix, and the retinal distance $D_{\rm ret}$ in the lower part. {\bf e)} Four pairs of stimulus clips (green and violet) are selected to demonstrate that $D_2$ (increasing top to bottom) and $D_{\rm ret}$ (increasing left to right) are not monotonically related.  } 
\label{f1}
\end{figure*}

To characterize the structure of neural distance in a large population, we recorded extracellularly the activity of $N=100$ retinal ganglion cells (RGCs) in the tiger salamander  using a multi-electrode array \cite{meister94,segev04}. The retina patch was presented 626 times with a $10\e{s}$ long segment of spatially uniform flicker with Gaussian distributed luminance drawn independently at $30\e{Hz}$ (Fig.~\ref{f1}a,b). Time was discretized into $T=961$ bins of $10\e{ms}$  and  the joint response of the  neurons $\sigma=\{\sigma_a\}$ was represented in each bin by a $N$-bit codeword, where $\sigma_a=1$ (0) denoted that the neuron $a=1,\dots,N$ spiked (did not spike) in that bin. Since direct sampling of the conditional distribution  $P(\sigma|s)$ is impractical for such a large population (even with hundreds of repeats, Fig.~\ref{f1}c), we inferred a stimulus-dependent maximum entropy (SDME) model for this data that predicts $P(\sigma|s)$ for each time bin, as we report in detail in Ref.~\cite{sdme}.

Since only differences in retinal responses can guide the organism's behavior, the biologically relevant distance between stimuli $s_1$ and $s_2$ must be a measure of similarity between their corresponding response distributions. We define the \emph{retinal distance} between the stimuli as the symmetrized Kullback-Leibler distance between the distributions of responses they elicit,
\begin{equation}
D_{\rm ret}(s_1,s_2)=D_{KL}^{\rm sym}(P(\sigma|s_1) , P(\sigma|s_2)) 
\end{equation}
where the symmetrized KL divergence is  $D_{KL}^{\rm sym}(p , q)= 0.5 \left(\sum_x (p(x) \log_2 p(x)/q(x) +q(x)\log_2 q(x)/p(x)\right)$ \cite{shannon}.  We choose this principled information-theoretic measure because it quantifies the difference between stimuli precisely to the extent that their response distributions are distinguishable \cite{shannon,butts}. 
Once constructed, the analysis of $D_{\rm ret}$ should help us uncover the fundamental aspects of the stimulus space, in particular, whether the distance between pairs of stimuli is determined by a small number of features or stimulus projections.

Using the SDME model for $P(\sigma|s)$ \footnote{SDME outperforms conditionally-independent models for $P(\sigma|s)$, but the low-dimensional qualitative structure of the stimulus space identified in this paper can be robustly reproduced with non-SDME models; see Supplementary Information.}, we computed the retinal distance, $D_{\rm ret}$, between every pair of stimulus clips in the experiment. Figure \ref{f1}d shows the profound difference between the Euclidean and retinal distance for all  the pairs in a  two-second interval of the stimulus. In particular, the same value of Euclidean distance can be obtained from very similar neural responses, or very different ones, as shown in Fig.~\ref{f1}e for selected pairs of stimulus clips.

\begin{figure}[b] 
\centering
\includegraphics[width=3in]{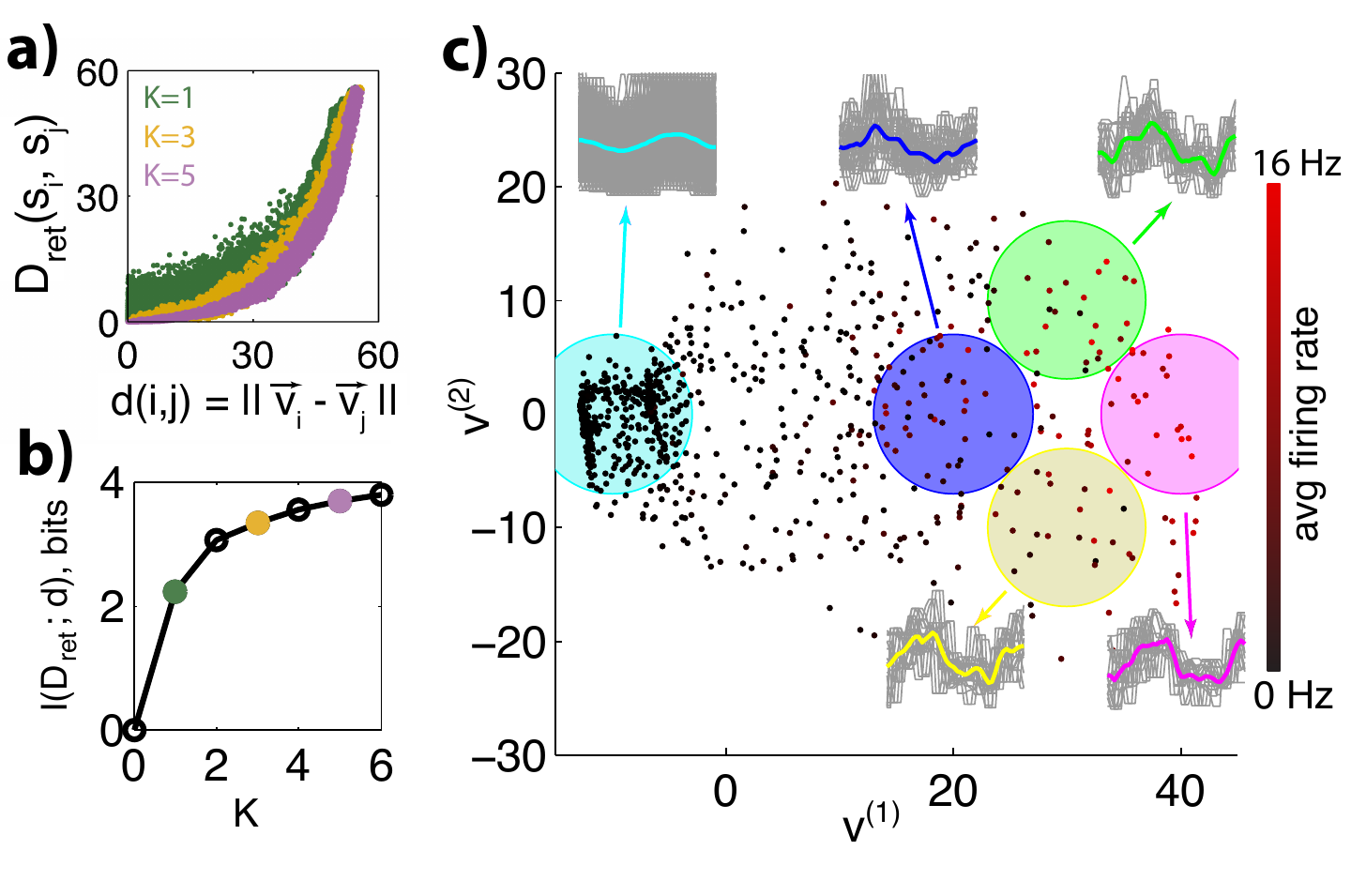} 
\caption{
MDS assigns to each stimulus clip $s_i$ a $K$-dimensional vector $\vec{v}_i=\{v^{(1)}_i,\dots,v^{(K)}_i\}$. 
{\bf a)}  Relationship between $D_{\rm ret}$  (one dot = one pair of stimuli) and the Euclidean norm in MDS space, $d(i,j)=||\vec{v}_i-\vec{v}_j||$, gets tighter with  $K$.
{\bf b)} Information $I$ between $D_{\rm ret}$ and MDS distance $d$ as a function of embedding dimension $K$. {\bf c)} For $K=2$, all $T=961$ stimulus clips are shown as points with MDS coordinates $(v^{(1)},v^{(2)})$; shade of red corresponds to mean population firing rate (scale at right). Five groups of stimuli are denoted by circles (gray lines = individual stimulus clips, color line = average). Equal distances in the plane correspond to equal $D_{\rm ret}$.} 
\label{f2}
\end{figure}

 \begin{figure*}[t] 
\centering
\includegraphics[width=6.5in]{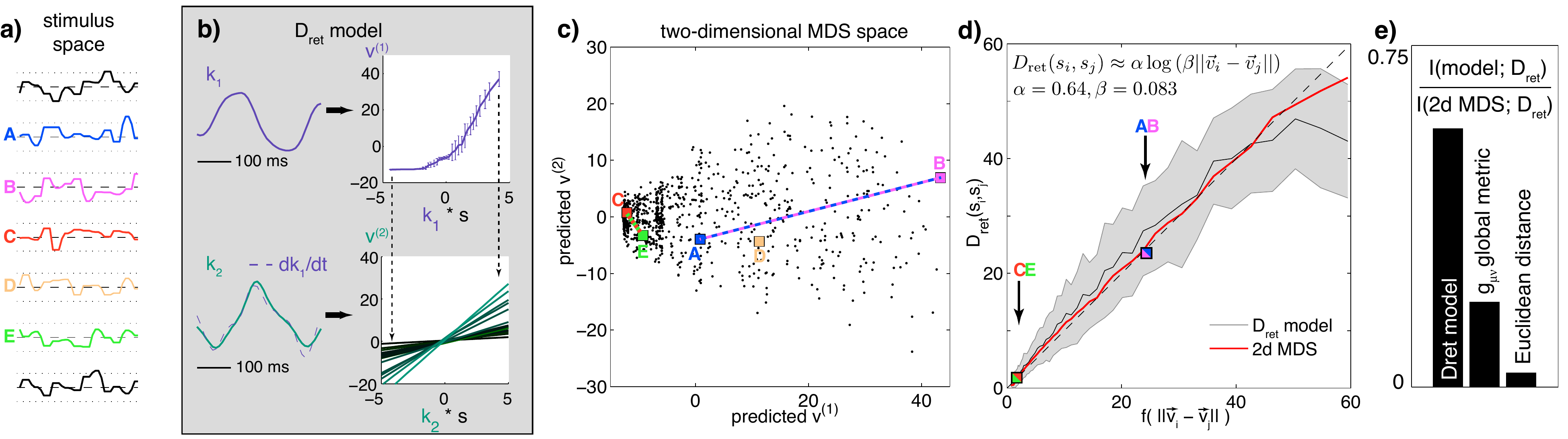} 
\caption{
{\bf a)} Sample clips from the stimulus space (5 highlighted in color). {\bf b)} A model for $D_{\rm ret}$  maps each stimulus clip $s$ to a point $(v^{(1)},v^{(2)})$ in the 2d MDS space. Predictions for $v^{(1,2)}$ are obtained by filtering the stimulus clip $s$ with a linear filter $k_{1,2}$  and passing the result through a pointwise nonlinearity (for $v^{(2)}$,  the transformation is linear with the slope that increases with $k_1\cdot s$, as shown). {\bf c)} All $T=961$ stimulus clips represented as points in a plane with  coordinates predicted by the model in b (colored squares = highlighted stimuli).  {\bf d)} Given predictions for coordinates $\vec{v}_i=(v^{(1)}_i,v^{(2)}_i)$, the $D_{\rm ret}$ for all ($\sim 4.6\E{5}$) pairs of clips $(s_i,s_j)$ can be predicted using a simple fitted relation in the inset (black line = mean binned model predictions vs true $D_{\rm ret}$, shaded area =  std of binned predictions). Red line shows $D_{\rm ret}$ computed from  true 2d MDS coordinates of Fig.~\ref{f2}c. Two highlighted  distances from c denoted by arrows. {\bf e)} The fraction of information about the true $D_{\rm ret}$ captured by the $D_{\rm ret}$ model in b, by the best-fit global metric $g_{\mu\nu}$, and by the Euclidean distance, normalized by the success of the $K=2$ MDS.
} 
\label{f3}
\end{figure*}

To further explore the structure of retinal distance, we used multidimensional scaling (MDS) to project  the distance matrix $D_{\rm ret}(s_i,s_j)$, between all pairs ($i,j=1,\dots,T$) of presented stimulus clips, into a low dimensional space \cite{mds}. This embedding technique  does not directly reveal the structure of the stimulus space, but can approximate its effective dimensionality. Every stimulus clip $s_i$  is assigned a $K$-dimensional point $\vec{v}_i$ in Euclidean space, such that 
\begin{equation}
D_{\rm ret}(s_i,s_j)\approx f(||\vec{v}_i-\vec{v}_j||), \label{mds}
\end{equation}
 with $f(\cdot)$ being a monotonic function. It is possible to find such accurate mappings for small $K$: Fig.~\ref{f2}a shows  the strong correspondence between low-dimensional MDS projections and the retinal distance, for different $K$ values. Fig.~\ref{f2}b summarizes the MDS performance at low orders in terms of the mutual information it captures about $D_{\rm ret}$; higher  information values correspond to a tighter relationship with less scatter. 
Figure \ref{f2}c shows the structure of stimulus space using MDS with $K=2$, which already captures most of the structure of the stimulus space. The first  coordinate of the MDS projection, $v^{(1)}$, is strongly correlated with the average firing rate in the population: high values correspond to  ``off-like'' stimuli, and small values  correspond to flat or ``on-like'' stimuli that do not drive the neural population well. The increased sensitivity to ``off-like'' features is consistent with the known prevalence of OFF-type cells in the salamander retina and in our dataset. Although  ``on-like'' stimuli differ substantially in their shapes and thus in their Euclidean distances $D_2$, they are largely indistinguishable for the retina. In contrast, groups of stimuli sharing the same $v^{(1)}$ coordinate (yellow and green) have a  similar  shape and evoke a similar population firing rate, yet are distinguishable because they differ along the second  coordinate $v^{(2)}$. To the retina, yellow and green groups of stimuli are as distinct from each other as the blue and magenta groups at constant $v^{(2)}$ but different  $v^{(1)}$. 

Figure~\ref{f3}  presents a model that predicts the mapping of an \emph{arbitrary} stimulus $s$ into $v^{(1)}$, $v^{(2)}$ and thus allows us to compute $D_{\rm ret}$. This model, obtained by using simple reverse correlation analysis \cite{schwartz06}, relies on two coupled linear-nonlinear transformations (Fig.~\ref{f3}a-d), and  identifies two dominant population-level stimulus features $k_1,k_2$: stimuli are distinguishable only insofar as their projections onto $k_1,k_2$ differ. The model  accurately predicts $D_{\rm ret}$ (Fig.~\ref{f3}e), establishing the relation of Eq.~(\ref{mds})  to be $d(i,j)=||\vec{v}_i-\vec{v}_j|| \approx \beta^{-1}\exp\{\alpha^{-1} D_{\rm ret}(s_i,s_j)\}$ (Fig.~\ref{f3}d). Interestingly, this relation, which we did not assume a priori, is exactly the kernel function used in several very successful applications of support vector machine (SVM) classification in machine learning, where one needs to distinguish between ``classes'' (here, stimulus clips) based on the distributions over ``features'' (here, neural responses) that they induce \cite{moreno03}. Our findings indicate that the neural population, much like single neurons, performs low-order dimensionality reduction on the incoming stimuli; however, unlike single neurons that can signal only a binary decision in every time bin, the population has access to $\sim 2^N$ states which can encode the variation along the relevant directions with greater precision. This view is consistent with the reported highly redundant code in the retina \cite{puchalla05}.

Given the failure of the Euclidean metric to predict stimulus similarity, and the success of the low-dimensional model, we asked whether a general quadratic form could  explain the retinal distance. We thus looked for an optimal matrix $g_{\mu\nu}$, such that $D_{\rm ret}(s_i,s_j) \approx \sum_{\mu\nu} (s_i^{(\mu)}-s_j^{(\mu)}) g_{\mu\nu} (s_i^{(\nu)}-s_j^{(\nu)})$, where $\mu, \nu$ range over all 40 components of stimulus clip vectors $s$. Using cross-validated least-squares fitting, we found that the optimal $g_{\mu\nu}$  substantially outperformed the Euclidean metric, yet still only captured $\sim 20\%$ of the structure in $D_{\rm ret}$ (Fig.~\ref{f3}e). 
The best-fitting matrix $g_{\mu\nu}$ has a simple structure that is captured by two eigenvectors, matching the  pair of population-level stimulus features, $k_1,k_2$, independently inferred in Fig.~\ref{f3}b.  Despite this, the best-fitting static $g_{\mu \nu}$ performs poorly: the eigenvalues corresponding to $k_1,k_2$ would  have to depend on the stimulus in order to approximate well our $D_{\rm ret}$ model.

\begin{figure} 
\centering
\includegraphics[width=3.6in]{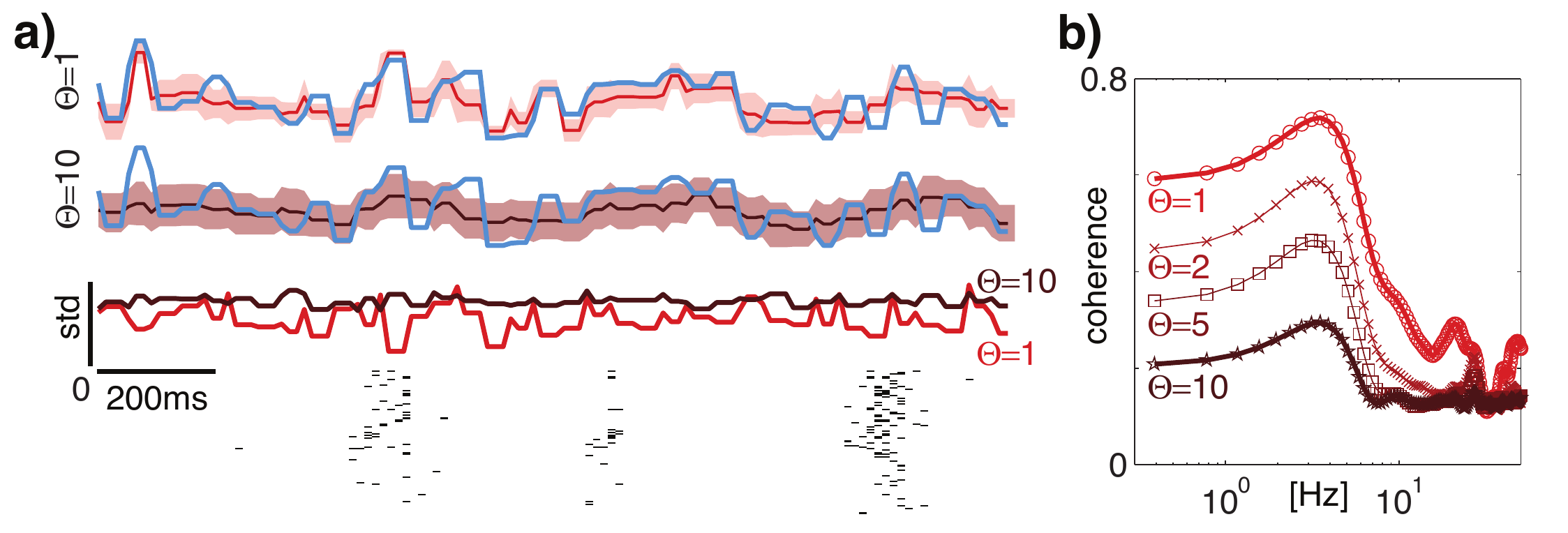} 
\caption{
{\bf a)} Two Monte-Carlo generated ensembles of long stimuli (ensemble mean = red; std = shade) where each clip of the generated stimulus is less than  $\Theta=1,10$ distant from the corresponding clip in the original stimulus (blue line).  Below:  std over the ensembles ($\Theta=1,10$), showing how ensembles get more tightly constrained at particular instants corresponding to high neural activity (raster, bottom) and steep stimulus changes. {\bf b)} The rise in coherence between the original stimulus and the simulated ensembles with decreasing $\Theta$.}
\label{f5}
\end{figure}

Our results carry important implications for stimulus decoding. The accuracy of our model for stimulus similarity enables us to create new stimuli that are similar to each other up to any desired distance.
We used Monte Carlo simulation to generate ensembles of full-length stimuli, such that \emph{each} clip from the generated stimulus is less than $\Theta$ distant (as measured by $D_{\rm ret}$) from the corresponding clip in the original stimulus displayed in the experiment (Fig.~\ref{f5}a).
Our analysis predicts that for small enough $\Theta$, all stimuli from such an ensemble are essentially indistinguishable to the retina. In contrast to Euclidean distance, which would allow the generated stimuli to fluctuate around the original waveform equally at every point in time for a given threshold $\Theta$, the retinal distance constrains the possible set of stimuli much more at certain times than at others, reflecting a preference of the retina for encoding specific features in the stimulus. This is in part due to compressive nonlinearities, which squeeze large segments of stimulus space with small feature overlap into a small volume as measured by retinal distance (Fig.~\ref{f2}c), while emphasizing stimulus differences with high feature overlap. Consequently, distance models (such as Euclidean distance) with constant metrics are incapable of capturing the characteristics of the retinal distance. We conjecture that decoding approaches based on minimizing standard distance measures (e.g. \cite{warland97}) may overly penalize deviations from aspects of the stimulus that are simply not  represented in the neural responses, while not emphasizing strongly enough the deviations from population-level stimulus features identified here. 

Understanding the neural code crucially depends on our ability to go beyond single-neuron spatio-temporal receptive fields, and identify how many, and which, are the population-level features in an interacting population. We introduced a novel, biologically relevant distance measure on the space of stimuli based on the activity of large populations of neurons. This approach extended the single-neuron notions of stimulus feature extraction to the neural population, determining how a high-dimensional input space is partitioned and encoded by population responses. Our work thus suggests a principled alternative to arbitrary norms (like the Euclidean norm) for stimulus similarity and decoding, generalizes previous attempts to construct metrics for particular input spaces from neural responses (e.g. \cite{curto,macke}), and complements existing work on the dual problem of constructing relevant spike-train distance measures \cite{victor,ya}.

The approach we  presented here will  be instrumental in the analysis of upcoming experiments, which allow the recording of large parts of sensory neural circuits, or even of all the cells encoding some parts of the sensory scene \cite{om}.  This approach can be immediately applied  to other sensory modalities, where it could signal---much as we have found here---that the ``neural metric''  deviates considerably from our intuitive notions of similarity. Moreover, it can be  extended to sensory domains where we lack any obvious notion of similarity, e.g. olfaction, for which there exists no natural distance between chemical stimuli \cite{haddad08}. More broadly, as the neural  metric is based on the spiking activity itself, this framework can be taken beyond sensory modalities, to study perceptual metrics as well (e.g. \cite{Freeman_11}) or  used to define neural-based distances for motor behavior that would be critical for neural prosthesis applications \cite{Velliste_08,Vargas-Irwin_10,Hatsopoulos+Donoghue}.

\end{document}